\documentclass[hyper, letterpaper,11pt,notoc]{JHEP3}
\usepackage{graphicx}
\usepackage{amssymb}
\usepackage{amsmath}
\usepackage{latexsym}
\usepackage[all]{xy}
\usepackage{feynmf}
\preprint{}

\title{Curvature as a remedy \\ or Discretizing gravity in warped dimensions}
\author{Jason Gallicchio and Itay Yavin \\ Jefferson Physical Laboratory, Harvard University, Cambridge, MA
02138  \\ E-mail: \email{jason@frank.harvard.edu}
\email{yavin@fas.harvard.edu} }

\abstract{The attempt to discretize gravity in flat space is foiled
by the appearance of strongly interacting long wave-length
longitudinal modes. In this paper we show how the introduction of
sites with different scales, or equivalently curvature in the bulk,
ameliorate all the problems encountered in flat space associated
with long wave-length modes. However, as one could expect, all such
problem resurface once the mode's wave-length is smaller than the
bulk curvature.}

\begin{document}
\begin{fmffile}{AdSpics}
\section{Introduction}
\label{sec: introduction}

It is an old dream in particle physics that the high-energy behavior of our theories is unified and regulated through
the disclosure of extra dimensions. An interesting twist to this dream was offered in recent years and is known as
\emph{De-construction}. It was shown that one can fabricate gauge theories in any dimension greater than 4, starting
with many copies of the same gauge group in 4 spacetime dimensions \cite{Arkani-Hamed:2001ca}. Any higher-dimensional
gauge theory has a dimension-full coupling and therefore grow strong in the UV. This construction, supplemented with an
UV completion for the non-linear sigma model describing the link fields (e.g. a linear sigma model, or dynamic symmetry
breaking), offers a complete description of the physics at all scales due to the asymptotically free nature of gauge
theories in 4-d. Shortly after, people attempted the de-construction of spin-2 particles
\cite{Alishahiha:2001nb,Bander:2001qk,Sugamoto:2001uk,Sugamoto:2002pk,Kan:2002rp}. The construction seemed like a
straight forward extension of the gauge theory case, and indeed at the linear level the low-energy physics behaves as
if a genuine extra dimension has emerged. At the same time, the minimal moose construction of
\cite{Arkani-Hamed:2002sp} elucidated the peculiar behavior of 4-d massive gravity and traced it back to the lack of a
kinetic term for the scalar longitudinal mode in the absence of coupling to the transverse part of the metric. This
realization immediately attaches a large question mark to any attempt of de-constructing gravity. The build up of an
extra dimension can always be thought of as involving many interacting massive spin-2 particles in 4-d. Therefore, any
problems one encounters with massive particles will inevitably hide in the background of any dimensional extension. And
indeed, as shown in \cite{Arkani-Hamed:2003vb}, the same problems of massive gravity are the origin of strong non-local
interactions of the longitudinal modes which render the low-energy degrees of freedom behaves nothing like an extra
dimension. While it is true that the linear analysis reveals the Kaluza-Klein spectrum, this is certainly not an extra
gravitational dimension.

Pinpointing the origin of the ailment is often the first step towards finding a cure. In \cite{Arkani-Hamed:2003vb},
the authors trace the problem to the lack of a kinetic term for the scalar longitudinal mode $\phi$ in the absence of
coupling to the transverse metric. In the notation of \cite{Arkani-Hamed:2003vb}, there is a mixing term of the
form,
\begin{equation}
 \mathcal{L} \supset \sum_j \frac{1}{a^2}\partial_\mu(h^{j+1} - h^j)\partial^\mu \phi^j
\end{equation}
where $i$ index the discrete extra dimension. After summation by parts and diagonalization of the mixing term with the
kinetic term for $h_{\mu\nu}$ one does generate a kinetic term,
\begin{equation}
\partial_\mu\frac{(\phi^{j+1}-\phi^{j})}{a^2}\partial^\mu \frac{(\phi^{j+1}-\phi^{j})}{a^2}
\end{equation}
Therefore it is $\psi = \partial_y \phi/a$ which is the propagating degree of freedom and not $\phi$ itself. This leads
to non-local interactions and strongly interacting long wave-length modes. Now that we understand the source of the
problem it is easier to conceive of a solution. Notice that had we had a scale factor in front of the mixing term
before integrating by parts there is a hope for a remedy. Starting with,
\begin{equation}
\mathcal{L} \supset f(y) \partial_\mu\partial_y h \partial^\mu \frac{\phi}{a}
\end{equation}
then when integrating by parts we will generate two terms,
\begin{equation}
\mathcal{L} \supset \partial_\mu h\left(f(y)\partial_y + \partial_yf(y)\right)\partial^\mu \phi
\end{equation}
If the healthy part $(\partial_yf(y))\phi$ dominates over $f(y)\partial_y\phi$ we would expect all the non-local
behavior to disappear and the dangerous amplitudes to be regulated. To get a better intuition into this condition it is
instructive to have a lattice interpretation of it. The condition then reads,
\begin{equation}
\label{eqn: healthiness condition}
\left|\frac{\phi_{j+1}-\phi_j}{\phi_j}\right| \ll \left|\frac{f_{j+1}-f_j}{f_j} \right|\hspace{5mm} or \hspace{5mm} \left|\frac{\phi_{j+1}}{\phi_j}\right| \ll \left|\frac{f_{j+1}}{f_j}\right|
\end{equation}
As long as the longitudinal modes vary much slower than the scale factor does, they will have a healthy kinetic term.
In particular if the number of sites is large (so the size of the space is large compared with the lattice spacing),
the low energy modes are well spread over the extra-dimension and change very little over one lattice-length. Condition
(\ref{eqn: healthiness condition}) is then easily satisfied. However, as the extra-dimension shrinks and becomes
comparable to the curvature, even the low energy modes will change more rapidly than the scale factor over one
lattice-length. This is the intuition which we will try to make precise in the rest of the paper.

The paper is organized as follows: In section \ref{sec: 2-site model} we analyze a 2-site model as a prelude for
constructing an extra dimension. Section \ref{Quadratic expansion for flat site metrics} contains the general
many-sites model. The flat extra dimension case and its demise are reviewed in section \ref{sec: flat extra dim}. We
then construct an extra dimension with an AdS profile in section \ref{sec: AdS extra dim} and investigate the long
wave-length modes behavior. We conclude in section \ref{sec: conclusions}.

\section{A 2-Site Model}
\label{sec: 2-site model}

\begin{figure}
\label{diag: two massive gauge}
\begin{equation*}
\xymatrix{
\ar@{-}[r]^{U_{1}(x)}_{f_1}& *+<10pt>[o][F]{A^\mu_1}& \ar@{-}[r]^{U_2(x)}_{f_2} & *+<10pt>[o][F] {A^\mu_2} & \ar@{=>}[r] & & \ar@{-}[r]^{U_{1}(x)}_{f_1}& *+<10pt>[o][F]{A^\mu_1}\ar@{-}[r]^{U_2(x)}_{f_2} & *+<10pt>[o][F] {A^\mu_2} \\ \ar@{=>}[r] && \ar@{-}[r]^{U_{1}(x)}_{f_1}& & *+<10pt>[o][F]{A^\mu_1} & \ar@{-}[r]^{U_2(x)}_{f_2}& & *+<10pt>[o][F] {A^\mu_2}
}
\end{equation*}
\caption{Two separate massive gauge theories can be linked to form a chain. Each of the chain's components is a healthy
theory and makes sense all by itself}
\end{figure}
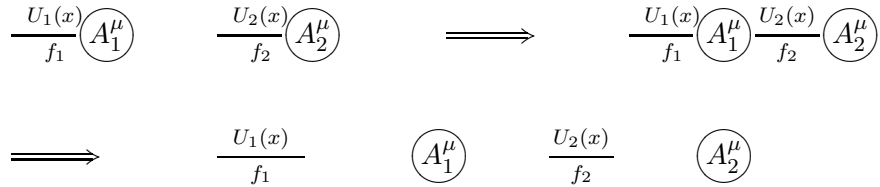

Before we plunge into a complicated many site model, it is very illuminating to consider a simple 2-site model.
Furthermore, it is instructive to contrast it with a 2-site gauge theory model for the differences sharply point to the
difficulties with gravity. To begin with, let's imagine we have two massive gauge theories, represented pictorially in
figure \ref{diag: two massive gauge}. The lagrangian describing the two theories is,
\begin{equation}
\mathcal{L} = \sum_{i=1,2} -\frac{1}{4g_i^2}F_{i,\mu\nu}F^{\mu\nu}_i + f_i^2 A_{i,\mu}A^\mu_i
\end{equation}
As expounded upon in \cite{Arkani-Hamed:2001ca} it is convenient to introduce the longitudinal modes as separate fields
by performing the broken gauge transformations and promoting those into a field. The lagrangian is then recast in the
form,
\begin{equation}
\mathcal{L} = \sum_{i=1,2} -\frac{1}{4g_i^2}F_{i,\mu\nu}F^{\mu\nu}_i + f_i^2 (D_\mu U_i)^\dagger(D^\mu U_i)
\end{equation}
where,
\begin{equation}
D^\mu U_i = \partial^\mu U_i + i U_i A^\mu_i
\end{equation}
As usual, the longitudinal polarizations are described by a non-linear $\sigma$-model with a pion decay constant $f_i$.
The second step in figure \ref{diag: two massive gauge} of linking the two theories is achieved by simply charging
$U_2$ with $A^\mu_1$ on the left so that, $D^\mu U_2 = \partial^\mu U_2 + i U_2 A^\mu_2 - i A^\mu_1 U_2$. The
importance of this two site example stems from two points it serves to illustrate regarding the prospects of adding
many more sites and building a dimension. First, in the limit where we decouple the transverse theories from the
longitudinal polarizations ($g_i\rightarrow 0$, keeping $f_i$'s fixed), each theory makes sense entirely by itself.
Second, glancing at the kinetic terms for the link fields $U_i$, we notice that $U_2$ has a linear coupling to
$A^\mu_1$ which in turn as a linear coupling to $U_1$. Adding many more sites, we might worry that distance link fields
will nonetheless couple strongly to each other through this mixing. Moreover, if we choose different decay constants
$f_i$, what ensures us that local excitations on one site will not sense the strong coupling regime on the other site
(which could be much lower). This is not unjustified, for as we will see below, this mixing is the origin of the
non-local interactions in the case of gravity. With gauge theories, however, this mixing is a misimpression and can be
removed by a simple gauge fixing choice (such as the $R_\xi$ gauges). This is why it is possible to deconstruct gauge
theories on $AdS_5$ for example, \cite{Randall:2002qr}.

Let's move on and consider a two-site model for gravity. We begin with two sites, each describing a massive spin-2
particle. We expand the metric about flat space, but allow for the possibility of different scales on the separate
sites,
\begin{equation}
g_{i,\mu\nu} = f_i^2(\eta_{\mu\nu} + h_{i,\mu\nu})
\end{equation}
For concreteness, we will take $f_1\ge f_2$ in all that follows. The action, which we will write in full glory only
once and use a schematic form henceforth, is given by,
\begin{eqnarray}
\mathcal{S} &=& \sum_{i=1,2} M^2f_i^2 \int d^4 x_i \left( \frac{1}{4}\partial _\mu  h_i^{\nu \rho} \partial ^\mu
h_{i,\nu \rho } - \frac{1}{4}\partial _\mu  h_i\partial ^\mu  h_i + \frac{1}{2}\partial _\mu  h_i\partial _\nu h_i^{\mu
\nu } - \frac{1}{2}\partial _\mu  h_i^{\nu \rho } \partial _\nu  h_{i,\rho}^\mu \right. \\\nonumber &+& \left.
\frac{m^2 f^2_i}{4}\left(h^2_i - h_{i,\mu\nu}h^{\mu\nu}_i\right) \right)
\end{eqnarray}
As usual, the Fierz-Pauli form of the mass term is chosen to guarantee the absence of any ghost-like polarizations.
Following the gauge theory example, it is convenient to introduce the longitudinal polarizations by performing the
broken diffeomorphism and promoting it into a field $h_{i,\mu\nu} \rightarrow h_{i,\mu\nu} - \nabla_\mu \pi_{i,\nu} -
\nabla_\nu \pi_{i,\mu}$. The lagrangian is now invariant under the transformations,
\begin{eqnarray}
h_{i,\mu\nu} &\rightarrow& h_{i,\mu\nu} - \nabla_\mu \epsilon_{i,\nu} - \nabla_\nu \epsilon_{i,\mu} \\
\pi_{i,\mu} &\rightarrow& \pi_{i,\mu} + \epsilon_{i,\mu}
\end{eqnarray}
The longitudinal polarizations can be further decomposed into a vector-like and scalar-like polarizations,
\begin{equation}
\pi_\mu = A_\mu + \partial_\mu \phi
\end{equation}
As shown in \cite{Arkani-Hamed:2001ca}, the vector modes form a simple $U(1)$ gauge theory which is perfectly healthy
and well-behaved. The interesting sector is the scalar $\phi$ and we will concentrate on it for the rest of the paper.
The mass term generates a kinetic term for $A_\mu$ in the usual form $F_{\mu\nu}^2$, however, there is no corresponding
kinetic term for $\phi$. It does however, couple linearly to the transverse modes and the relevant term in the
lagrangian is,
\begin{eqnarray}
\mathcal{L} &\supset& \sum_{i=1,2} M^2m^2f_i^4 \left( h_{i,\mu\nu}-\partial_\mu\partial_\nu\phi_i\right)\left(
\eta^{\mu\nu}\eta^{\alpha\beta}- \eta^{\mu\alpha}\eta^{\nu\beta}\right) \left(
h_{i,\alpha\beta}-\partial_\alpha\partial_\beta\phi_i\right) \\\nonumber &\supset& \sum_{i=1,2} M^2m^2f_i^4 \left(
h_i\Box\phi_i - h_{i,\mu\nu}\partial^\mu\partial^\nu\phi_i\right)
\end{eqnarray}
We are omitting higher-order terms and in particular $(\Box\phi)^3$ to avoid clutter. Those are of course important and
lead to the growing amplitudes which mark the breakdown of the effective theory. We will spell all of these in details
later on, but for now it is only their existence which is important and not the precise form.

As clearly illustrated in \cite{Arkani-Hamed:2001ca} the fact that $\phi$ doesn't have a kinetic term of its own is the
origin of all the peculiarities of massive gravity and we wish to expand upon this point no more.
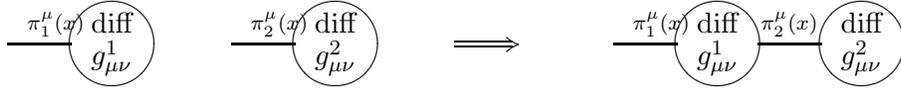
\begin{figure}
\label{diag: two massive gravity model}
\begin{equation*}
\xymatrix{ \ar@{-}[r]^{\pi^\mu_{1}(x)}& *+<15pt>[o][F]\txt{diff \\ $g_{\mu\nu}^1$}& \ar@{-}[r]^{\pi^\mu_2(x)} &
*+<15pt>[o][F]\txt{diff \\ $g_{\mu\nu}^2$} & \ar@{=>}[r]&& \ar@{-}[r]^{\pi^\mu_{1}(x)}& *+<15pt>[o][F]\txt{diff \\
$g_{\mu\nu}^1$} \ar@{-}[r]^{\pi^\mu_2(x)} & *+<15pt>[o][F]\txt{diff \\ $g_{\mu\nu}^2$} }
\end{equation*}
\caption{Two separate massive gravity theories linked together}
\end{figure}
At this stage we would like to link the two sites and investigate the behavior of the resulting theory. As in the gauge
case this can be done by charging the pions of the second site under the diffeomorphism of the first site, and we
relegate the details of this step into section \ref{sec:many site model}. The part of the lagrangian which is of
interest to us is,
\begin{eqnarray}
\label{eqn: h-phi mixing term unrescaled} \mathcal{L}&\supset& M^2m^2f_1^2\left( h_1\Box\phi_1 -
h_{1,\mu\nu}\partial^\mu\partial^\nu\phi_1\right) \\\nonumber &+& M^2m^2f_2^2 \left( (h_2-h_1)\Box\phi_2 -
(h_{2,\mu\nu}-h_{1,\mu\nu})\partial^\mu\partial^\nu\phi_2\right)
\end{eqnarray}
Unlike the gauge theory, this mixing between the scalar and transverse modes cannot be eliminated by any gauge choice.
In order to diagonalize the kinetic term we must perform a Weyl transformation as in
\cite{Arkani-Hamed:2002sp,Arkani-Hamed:2003vb}. However, before we proceed any further we would like to formulate 3
questions on which the prospects of extending this two-site model into a chain hinge:
\begin{enumerate}
\item One might worry that coupling site-1 to site-2, which has a lower scale, will render the strong coupling scale on
site-1 lower as well. In particular, an experimentalist on site-1 (a source on site-1) might discover the strong
coupling scale to be much lower than expected. If this is indeed the case, we have no right to claim that a many-site
model mimics AdS in any way. We will show this not to be the case.
\item Can we understand what goes wrong when $f_1=f_2$? Extrapolating to a many-site model this is the flat extra dimension case.
\item Can we understand the behavior of long wave-length modes in the many-site model, through this two-site model?
\end{enumerate}
To answer the first question we will show that both the strong coupling scale, encountered when performing scattering
experiments, and the Vainshtein radius where the effective theory breaks down around heavy sources, are as expected for
scales on site-1. This is actually fairly trivial to see. We begin by canonically normalizing our fields,
\begin{eqnarray}
h_{i,\mu\nu} &\rightarrow& Mf_i h_{i,\mu\nu} \\
\phi_i &\rightarrow& Mm^2f_i^3 \phi_i
\end{eqnarray}
The mixing term (\ref{eqn: h-phi mixing term unrescaled}) takes the form (to avoid clutter we abbreviate the
Fierz-Pauli form and present the trace part only),
\begin{eqnarray}
\mathcal{L} &\supset& h_1\Box \phi_1 + (h_2 - \epsilon h_1)\Box \phi_2  \\\nonumber &=& h_1\Box(\phi_1 - \epsilon
\phi_2) + h_2\Box\phi_2 \hspace{10mm} \epsilon = \frac{f_2}{f_1}
\end{eqnarray}
and the source terms for each site are,
\begin{equation}
\mathcal{L}_{source} = \frac{1}{f_1M}T^{\mu\nu}_1 h_{1\mu\nu} + \frac{1}{f_2M} T^{\mu\nu}_2 h_{2\mu\nu}
\end{equation}
The mixing between the transverse $h_i$ and the longitudinal $\phi_i$ is eliminated by a Weyl re-scaling,
\begin{eqnarray}
h_{1,\mu\nu} &\rightarrow& h_{1,\mu\nu} - \eta_{\mu\nu}(\phi_1-\epsilon \phi_2) \\
h_{2,\mu\nu} &\rightarrow& h_{2,\mu\nu} - \eta_{\mu\nu} \phi_2
\end{eqnarray}
The resulting longitudinal modes theory is given schematically by,
\begin{eqnarray}
\label{eqn: resulting long th}
 \mathcal{L} &\supset& (\phi_1-\epsilon \phi_2)\Box (\phi_1-\epsilon \phi_2) + \phi_2\Box
\phi_2 + \frac{1}{f_1M}T_1 (\phi_1-\epsilon\phi_2) + \frac{1}{f_2M}T_2\phi_2 \\\nonumber &+&
\frac{1}{f_1^5m^4M}(\Box\phi_1)^3+\frac{1}{f_2^5m^4M}(\Box\phi_2)^3 + \ldots
\end{eqnarray}
It is clear that it is $\phi_1' = \phi_1-\epsilon\phi_2$ and $\phi_2' = \phi_2$ which propagate and couple to the
sources on site-1 and site-2, respectively. This is important. Had we started off with $\phi_1$ coupling directly to
the source on site-1, we would end up with $\phi_2'$ having direct coupling to site-1 and that would lower all the
scales. The interaction terms (such as the trilinear coupling in equation \ref{eqn: resulting long th}), will now
involve couplings between $\phi_1'$ and $\phi_2'$, but those are all suppressed by powers of $\epsilon$. A scattering
experiment on site-1 will be dominated by the trilinear coupling $(\Box\phi_1')^3/(f_1m^4M)$ which give rise to the
usual divergent amplitude,

\begin{equation}
{\cal A} \quad=\quad
%
% phi1 phi1 -> phi1 phi1
%
\parbox{30mm}{
\begin{fmfgraph*}(30,30)
\fmfleft{p1,p2}
 \fmfright{q1,q2}
  \fmf{dashes}{p1,pl}
   \fmf{dashes}{p2,pl}
   \fmf{dashes}{pl,pr}
   \fmf{dashes}{pr,q1}
   \fmf{dashes}{pr,q2}
   \fmfv{l=$\phi_1'$,l.a=120,l.d=.03w}{p1}
\fmfv{l=$\phi_1'$,l.a=-120,l.d=.03w}{p2} \fmfv{l=$\phi_1'$,l.a=60,l.d=.03w}{q1} \fmfv{l=$\phi_1'$,l.a=-60,l.d=.03w}{q2}
\end{fmfgraph*} } \quad  \sim \quad
\frac{E^{10}}{\Lambda^{10}_1} \label{diag: scalarscattering}
\end{equation}

where $\Lambda_1 = f_1(m^4M)^{1/5}$ which correspond to the local scale on site-1. But, there are other types of
experiments one can perform. As Vainshtein showed \cite{Vainshtein:1972sx}, the effective theory describing massive
gravity breaks down around heavy sources at a much lower scale than $\lambda_1$. Since the longitudinal mode couples to
the trace of the energy-momentum tensor, a massive source $M_{s1}$ on site-1 sets up a potential that at linear level
goes as,
\begin{equation}
%
% 1->1
%
V^{(1)}(r) \sim \hspace{10mm}\parbox{30mm}{
\begin{fmfgraph*}(30,20)
\fmfleft{pl} \fmfright{pr} \fmf{dashes}{pl,pr}
\fmfv{decoration.shape=circle,decoration.filled=hatched,decoration.size=.07width}{pl}
\end{fmfgraph*} } \quad\quad
\sim \frac{M_{s1}}{M}\frac{1}{r}
\end{equation}
The leading correction comes from the trilinear coupling and to first order it contributes,
\begin{equation}
%
% 2->1
%
V^{(2)}(r) \sim \hspace{10mm}
\parbox{30mm}{
\begin{fmfgraph*}(30,20)
\fmfleft{p1,p2} \fmfright{pr} \fmf{dashes}{p1,pl} \fmf{dashes}{p2,pl} \fmf{dashes}{pl,pr}
\fmfv{decoration.shape=circle,decoration.filled=hatched,decoration.size=.07width}{p1}
\fmfv{decoration.shape=circle,decoration.filled=hatched,decoration.size=.07width}{p2}
\end{fmfgraph*} } \quad\quad
\sim \left(\frac{M_{s1}}{f_1M}\right)^2\frac{1}{\Lambda^5_1}\frac{1}{r^6}
\end{equation}
The radius at which this correction becomes comparable to the linear contribution is simply,
\begin{equation}
r_V = \left(\frac{M_{s1}}{f_1M}\right)^{1/5} \frac{1}{\Lambda_1}
\end{equation}
Which, for heavy sources, is much lower than the strong coupling scale $\Lambda_1$. This is the Vainshtein result.
Notice that all the scales are as expected for an observer located on site-1. One is justified in worrying that heavy
sources on site-2 might lower the Vainshtein scale through mixing terms such as
$\epsilon(\Box\phi_1')^2(\Box\phi_2')/(f_1m^4M)$. This is possible, and could happen when considering second order
contributions. However, it is fairly easy to see that as long as,
\begin{equation}
\frac{M_{s2}}{M_{s1}} \ll \frac{f_2}{f_1}
\end{equation}
all such contributions are suppressed. This is a more precise version of condition \ref{eqn: healthiness condition},
involving a bound on the sources, rather than field amplitudes. These considerations lend credence to the possibility
of extending this two-site model into an extra dimension with a varying scale factor.

 It is also evident that mixing is maximal when $\epsilon=1$ which spells disaster for a flat chain. Since $\phi_2$
mixes maximally with $h_1$ which in turn mixes with $\phi_1$ which mixes with $h_0$ etc. very distant sites will be
strongly correlated and non-locality is sure to emerge. It is simple enough to see the form of the mixing terms when we
consider many-sites model,
\begin{equation}
\mathcal{L} \supset h_1\Box(\phi_1 - \epsilon \phi_2) + h_2\Box(\phi_2 - \epsilon\phi_3) + \ldots = \sum_i
h_i\Box(\phi_i - \epsilon\phi_{i+1}) + \frac{1}{f_i^5 m^4 M}(\Box\phi_i)^3
\end{equation}
Where, for the purpose of illustration we have included the trilinear coupling which leads to divergent amplitudes. The
important point is that while it is $\phi_i$ which interacts the combination that receives a kinetic term is,
\begin{equation}
\psi_i = A_{ij} \phi_j \hspace{10mm} A =\left(%
\begin{array}{ccccc}
  \ddots & \ddots & \hfill & \hfill & \hfill \\
1 & -\epsilon & 0 & \hfill & \hfill \\
0 & 1 & -\epsilon & 0 & \hfill \\
\hfill & 0 & 1 & -\epsilon & 0 \\
\hfill & \hfill & \hfill & \ddots & \ddots \\\end{array}%
\right)
\end{equation}
Since it is $\phi_j$ which appears in interactions we better express it in terms of $\psi_i$ through the inverse of $A$,
\begin{equation}
A^{-1} = \left(%
\begin{array}{ccccc}
  \ddots & \ddots & \hfill & \hfill & \hfill \\
1 & \epsilon & \epsilon^2 & \epsilon^3 & \epsilon^4 \\
0 & 1 & \epsilon & \epsilon^2 & \epsilon^3 \\
\hfill & 0 & 1 & \epsilon & \epsilon^2 \\
\hfill & \hfill & \hfill & \ddots & \ddots \\\end{array}%
\right)
\end{equation}
Which makes the non-local nature of the interactions clear when $\epsilon=1$. In contrast when $\epsilon$ is small,
long wave-length modes where neighboring sites are correlated have perfectly local interactions. The rest of this paper
is just a rephrasing of these arguments in the continuum form.

\section{Many Sites Model}
\label{sec:many site model}

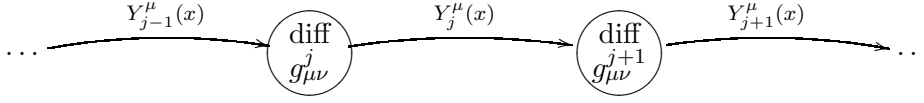
\begin{figure}
\label{diag: flat moose}
\begin{equation*}
\xymatrix{ {\ldots} \ar@/^/[rrr]^{Y^\mu_{j-1}(x)}&& & *+<15pt>[o][F]\txt{diff \\ $g_{\mu\nu}^j$}
\ar@/^/[rrr]^{Y^\mu_j(x)} &&& *+<15pt>[o][F]\txt{diff \\ $g_{\mu\nu}^{j+1}$}  \ar@/^/[rrr]^{Y^\mu_{j+1}(x)} &&&
{\ldots} }
\end{equation*}
\caption{A flat moose}
\end{figure}

A theory describing many interacting spin-2 particles may be represented graphically using the moose notation shown in
figure (\ref{diag: flat moose}), as in \cite{Arkani-Hamed:2003vb}. Each of the sites is endowed with its own metric
$g_{\mu \nu}$ and accompanying diffeomorphism. Diffeomorphism-invariant interaction terms require link fields
$Y_j^\mu(x_j)$ that can be thought of as an embedding of site $j$ onto site $j+1$. In other words $Y_j^\mu(x_j)$ takes
a point in site $j$ and maps it to a point $Y_j^\mu$ in site $j+1$. It transforms as a coordinate on site $j+1$ and is
a useful object since it acts as a comparator allowing us to compare fields on site $j+1$ to those on site $j$.
Applying it to the metric on site $j+1$, we can map it to a neighboring site $j$.
\begin{equation}
\label{eqn:induced metric}
G^{j+1}_{\mu\nu} = \frac{\partial Y_j^\alpha}{\partial x_j^\mu}\frac{\partial Y_j^\beta}{\partial x_j^\nu} g^{j+1}_{\alpha\beta}
\end{equation}
This is simply the induced metric and it transforms as a scalar under site $j+1$'s diffeomorphism and as a rank-2
tensor under site $j$'s diffeomorphism.

Using the link field we can now build interactions between arbitrary sites respecting each one's diffeomorphism. If our
goal is to build a real discretized dimension, large differences between the field value at neighboring points should
cost energy. This is the purpose of the gradient term in field theory. We need to subtract from the Lagrangian the
discrete analog of the gradient kinetic term. However, the different sites can have completely different scales. We
would not want to pay energy for simply a difference in scales among the sites, which motivates the following
comparison term between the sites,
\begin{equation}
\label{eqn:difference in metrics}
\Delta _{\mu \nu }^j  = \frac{1}
{a}\left( {\left( {\frac{{f_j }}
{{f_{j + 1} }}} \right)^2 G_{\mu \nu }^{j + 1}  - g_{\mu \nu }^j } \right).
\end{equation}
where $f_j$ is the scale factor on site $j$. If one is uncomfortable with the appearance of the ratio of scales, this
can be relegated to the definition of the mapping $Y_j^\mu$.

In the case of flat space the scale doesn't vary among the sites and we simply recover the case considered previously.
The two ways to combine this object into a scalar quadratic in the metric for use in the lagrangian are clearly $\Delta
^2$ and $\Delta _{\mu \nu } \Delta ^{\mu \nu }$. The usual Fierz-Pauli form is required to avoid the propagation of a
ghost (see \cite{Arkani-Hamed:2003vb}).
\begin{eqnarray}
\label{eqn:Ssite Slink}
\mathcal{S} &=& \mathcal{S}_{site}+\mathcal{S}_{link} \\
\label{eqn:Action with Delta}
&=& \int {d^4 x} \sum\limits_j {\sqrt { - g_{\mu \nu }^j } M^2 \left\{ { -R[g_{\mu \nu }^j ] + \frac{1}
{4}\left( {\Delta _j ^2  - \Delta _{\mu \nu }^j \Delta _j^{\mu \nu } } \right)} \right\}}
\end{eqnarray}

If anything is to go wrong with this model, experience from the flat profile and massive gravity cases show it would be
the longitudinal modes. To analyze the behavior of the longitudinal modes, unitary gauge is inappropriate and we
consider small fluctuations about the identity mapping,
\begin{equation}
\label{eqn: general gauge}
Y_j^\mu = x_j^\mu + \pi_j^\mu(x)
\end{equation}
The induced metric (\ref{eqn:induced metric}) to linear order in $\pi_j^\mu(x)$ is given by
\begin{equation}
\label{eqn: induced metric general gauge}
G^{j+1}_{\mu\nu} = g^{j+1}_{\mu\nu} + \nabla^{j+1}_\mu \pi^j_\nu + \nabla^{j+1}_\nu \pi^j_\mu + \mathcal{O}(\pi^2) \cdots
\end{equation}
where covariant derivatives and lowering of the indices on $\pi^j_\mu$ use $g^{j+1}_{\mu\nu}$ with any profile factor
included. This is the form of an infinitesimal coordinate transformation $x'=x+\epsilon(x)$ where $g'_{\mu\nu}=
g_{\mu\nu} + \nabla_\mu \epsilon_\nu + \nabla_\nu \epsilon_\mu$ and is what maintains general covariance on each site
since $\pi'_{\mu}= \pi_\mu-\epsilon_\mu$. It is also convenient to introduce a $U(1)$ gauge redundancy and decompose
the longitudinal modes into a vector and scalar parts,
\begin{equation}
\label{eqn: decomposing the longitudinal modes}
\pi_\mu = A_\mu + \nabla_\mu \phi
\end{equation}
From the structure of $G^{j+1}_{\mu\nu}$ or from the transformation properties we enforce on our fields it is clear
$\pi_\mu$ only appears with derivatives and has no mass term. Therefore $\phi$ can only appear with two derivatives and
has no kinetic term. As remarked above, the Fierz-Pauli choice guarantees the absence of any term of the form
$(\Box\phi)^2$ which will inevitably lead to a propagating ghost.

\section{The general profile lagrangian}
\label{Quadratic expansion for flat site metrics}
Let's now consider a theory space where the different sites can have different scales and expand each site's metric about flat space,
\begin{equation}
\label{eqn: site metric}
g^j_{\mu\nu} = f_j^2 \eta_{\mu\nu} + f_j h^j_{\mu\nu}
\end{equation}
where $f_j$ is the local scale factor, which is $e^{-k(ja)}$ to get 5D AdS. The power of $f_j$ in front of
$h^j_{\mu\nu}$ is of course arbitrary and is chosen so that the resulting kinetic 4D term for $h^j$ will have no
factors of $f$.  Everything with 4D $\mu\nu$ indices is raised with the common background metric, in this case
$\eta_{\mu\nu}$.  The comparison term becomes
\begin{equation}
\Delta _{\mu \nu } ^j  = \frac{1}
{a}\left( {\frac{{f_j ^2 }}
{{f_{j + 1} ^2 }}G_{\mu \nu }^{j + 1}  - g_{\mu \nu }^j } \right)
\end{equation}

We want to keep careful track of our factors of $f$ that appear in $\Delta_{\mu\nu}$,  $\sqrt { - g_{\mu \nu } }$, and
raising and lowering indices, so from here down, we'll use the background site metric $\bar g_{\mu \nu } (=
\eta_{\mu\nu})$ to raise and lower the indices.
\begin{equation}
G_{\mu \nu }^{j + 1}  = f_{j + 1}^2 \left(  \bar g_{\mu \nu } + \bar \nabla _\mu  \pi _\nu   + \bar \nabla _\nu  \pi
_\mu  \right)  + f_{j + 1} h_{\mu \nu }^{j + 1}  + \mathcal{O}(\pi^2) \cdots
\end{equation}

It is a notational monstrosity to work with the discrete index so we convert to the continuum langauge where
\begin{equation}
a \sum_j \rightarrow \int dy \hspace{10mm} \frac{M^2}{a} \rightarrow M^3_{5D} \hspace{10mm} \frac{h_{j+1}-h_{j}}{a} \rightarrow \partial_y h(y)
\end{equation}
The differences along the chain are regular (as opposed to covariant) derivatives and integration by parts will
explicitly deal with the factors of $f(y)$ that come from $\sqrt { - g_{\mu \nu } }$ even though in the continuum
language the covariant derivatives in the direction of the extra dimension made this task easy.  The continuum langauge
is more of a pneumonic than a formal limit since we want to keep $a$ finite. With these replacements we have,

\begin{equation}
\Delta _{\mu \nu } (y) = f^2 \left( {\left( {\partial _y \frac{{h_{\mu \nu } }}
{f}} \right) + \frac{2}
{a} \bar \nabla _\mu  \bar \nabla _\nu  \phi } \right)
\end{equation}

Taking the action (\ref{eqn:Action with Delta}) and the field definitions (\ref{eqn: decomposing the longitudinal
modes}-\ref{eqn: induced metric general gauge}), expanding to quadratic order, and ignoring the $A_\mu$ vector modes,
\begin{eqnarray}
\mathcal{S} =
\int {d^4 xdy\sqrt { - \bar g_{\mu \nu } } M_{5D} ^3 } && \left\{
\frac{1}{8}\partial _\mu  h^{\nu \rho } \partial ^\mu  h_{\nu \rho }
- \frac{1}{8}\partial _\mu  h\partial ^\mu  h
+ \frac{1}{4}\partial _\mu  h\partial _\nu  h^{\mu \nu }
- \frac{1}{4}\partial _\mu  h^{\nu \rho } \partial _\nu  h_\rho  ^\mu \right.  \hfill \\ \nonumber
&& \left. + \frac{{f^4 }}{8}\left[ {\left( {\partial _y \frac{h}{f}} \right)^2
- \left( {\partial _y \frac{{h^{\mu \nu } }}{f}} \right)\left( {\partial _y \frac{{h_{\mu \nu } }}{f}} \right)} \right]   \right.   \hfill \\ \nonumber
&& \left. + \frac{{f^4 }}{{2a}}\left[ {\left( {\partial ^2 \phi } \right)\left( {\partial _y \frac{h}{f}} \right)
- \left( {\partial _\mu  \partial _\nu  \phi } \right)\left( {\partial _y \frac{{h^{\mu \nu } }}{f}} \right)} \right]
 + \mathcal{O}(3)\cdots\right\}.
\end{eqnarray}
The quadratic piece in $h_{\mu\nu}$ is precisely what one obtains when starting with the 5-d AdS lagrangian for the
graviton. This last line's factors of $f$ came from a careful counting in the hopping term, explicitly $\sqrt { -
g_{\mu \nu } } \Delta _{\mu \nu } \left( {g^{\mu \nu } g^{\sigma \rho }  - g^{\mu \sigma } g^{\nu \rho } }
\right)\Delta _{\sigma \rho }$. It is a kinetic mixing term between $h_{\mu\nu}$ and $\phi$, and to remove it, we
change variables
\begin{equation}
\label{eqn:variable change}
  h_{\mu \nu }  \to h_{\mu \nu }  + \psi  \hspace{10mm}
  \frac{1}
{{af}}\partial _y \left( {f^4 \phi } \right) \to \psi
\end{equation}

This is a linearized Weyl re-scaling for $h_{\mu\nu}$ and gives $\psi$ a healthy kinetic term making it a propagating
field. The first two lines involving only $h_{\mu\nu}$ don't change, and the action becomes
\begin{equation}
\label{eqn:action with h and psi}
\mathcal{S} = \int {d^4 xdy M_{5D}^3} \left\{ {\mathcal{L}(h) + \frac{3}
{4}\partial _\mu  \psi \partial ^\mu  \psi  + \frac{3}
{2}f^4 \left( {\partial _y \frac{\psi }
{f}} \right)^2  - \frac{3}
{4}f^4 \left( {\partial _y \frac{h}
{f}} \right)\left( {\partial _y \frac{\psi }
{f}} \right) +  \mathcal{O}(3)\cdots } \right\}
\end{equation}

Since it is $\phi$ and not $\psi$ which forms the higher-order interactions, it is instructive to invert the operator.
\begin{equation}
\label{eqn: phi in terms of psi}
\phi(y) = \frac{a}{f(y)^4}\int_y f(y') \psi(y') dy'
\end{equation}

Obviously, this field redefinition has no effect at the linear level (like any invertible field redefinition), which is
why naively it seems like an extra dimension as emerged.  The problematic low-energy non-local interactions found in
\cite{Arkani-Hamed:2003vb} come about because $\phi = a\int \psi dy$, and it is $\phi$, not $\psi$ that comes into the
higher-order interaction terms in the expansion (\ref{eqn: induced metric general gauge}).

We want to count the factors of $a$ and $f$ that appear in the higher order terms.  There's a factor of $1/a^2$ in
front of the terms that come from $\Delta ^2  - \Delta _{\mu \nu } \Delta^{\mu \nu }$.  Each time you create a
derivative, you absorb a factor of $a$, but $\Box\phi$ never absorbs derivatives, and since it's massless, it always
comes with a $1/a$. The whole term comes with $f^4$ purely from the $\sqrt { - \bar g_{\mu \nu } }$ and each
$h_{\mu\nu}$ and $\psi$ needs one in it's denominator.

\section{Problems with flat profile}
\label{sec: flat extra dim}
 A flat profile has $f(y)=1$ and as shown in \cite{Arkani-Hamed:2003vb,Schwartz:2003vj}, the
most dangerous interaction is the cubic coupling $(\Box\phi)^3/a^2$. Let us briefly review the failure of the flat
moose to generate an extra dimension. If we consider a long wave-length mode $\psi_R(x)$ in an the extra dimension of
size $R$ and using equation (\ref{eqn: phi in terms of psi}) to find $\phi$, it is straightforward to derive the
effective action for $\psi_R(x)$ from (\ref{eqn:action with h and psi}),
\begin{equation}
\int d^4x M_{5D}^3 R \left( -\psi_R\Box\psi_R - R^3 a(\Box \psi_R)^3\right)
\end{equation}
Inspecting the $\phi\phi\rightarrow \phi\phi$ scattering in flat space one finds the same growth with energy as in
\ref{diag: scalarscattering}. This scattering amplitude grows strongly with energy and the effective theory breaks down
at energies $E\sim\Lambda$, where,
\begin{equation}
\label{eqn: flat cutoff}
\Lambda^{10} = \frac{M^3_{5D}}{a^2R^5}
\end{equation}
It is tempting to try and take the limit $a\rightarrow 0$, but this is impossible. The heaviest KK mode of the lattice
oscillates with wavelength of order $a$ and has a mass $m_n \propto 1/a$, which grows faster than the cutoff as
$a\rightarrow 0$. For the effective theory to make any sense, the cutoff $\Lambda$ needs to be above the heaviest mode.
Notice also that the size of the mode $R$ is in the denominator of equation (\ref{eqn: flat cutoff}) and therefore it
is the largest modes which are most dangerous. In particular the cutoff of the theory is much lower than the genuine
5-d theory $M_{5D}$. The low energy degrees of freedom are simply not the longitudinal modes of a genuine extra
dimension as their strongly non-local interactions indicate.

\section{AdS Profile}
\label{sec: AdS extra dim}

The profile that generates a 5D AdS background is $f(y) = exp(-k y)$ for which equation (\ref{eqn: phi in terms of
psi}) becomes
\begin{equation}
\label{eqn:AdS phi in terms of psi} \phi(y) = a e^{4 k y}\int_y e^{-k y} \psi(y') dy' = ae^{4ky} \frac{1}{k}e^{-ky}
\left(\psi(y) + \frac{1}{k}\frac{d\psi}{dy} + \frac{1}{k^2}\frac{d^2\psi}{dy^2} + \ldots\right)
\end{equation}
Whereas in the flat moose case the integral over the long wave-length modes scaled as the size of the modes, the curved
moose regulates the integral through the introduction of the curved profile. It is clear that any mode with a
wave-length much bigger compared to the radius of curvature $1/k$ is such that,
\begin{equation}
\phi(x,y) = \frac{a e^{3ky}\psi(x,y)}{k}
\end{equation}
This already looks promising. The interaction is no longer spread over the entire space, but is rather local. We have
been a little sloppy regarding the way this integral scales. To illustrate this point consider the integral over
$exp(-ky)$ and recall that this integral is just an approximation for a discrete sum,
\begin{equation}
\int e^{-ky} dy \approx \sum_i a e^{-ika} = \frac{a}{1-e^{-ka}}
\end{equation}
Indeed, when $ka$ is not too large $exp(-ka) \approx 1-ka$ and the integral scales as $1/k$. However, when $ka\gg 1$,
the integral scales as the lattice spacing $a$. This is simply the statement that you can only be as sensitive as your
lattice spacing. With this in mind, let's then try to calculate the same amplitude as before for the case of an AdS
profile function $f(y) = exp(-ky)$. The troublesome interaction term is $f^4(\Box\phi)^3/a^2$. If we consider a mode of
size $R$ which extends all the way to $y$ in the bulk, then using equation (\ref{eqn: phi in terms of psi}) once more
we can write the effective action for a long wave-length mode compared with the curvature $1/k$,
\begin{equation}
\int d^4 x M_{5D}^3\left(R\psi_R\Box\psi_R + \frac{a}{k^4}e^{5 k y} (\Box\psi_R)^3\right)
\end{equation}
It is easy to compute the same $\phi\phi\rightarrow\phi\phi$ scattering amplitude as before. The same growth with
energy is present but the cutoff is different,
\begin{eqnarray}
\label{eqn: AdS Cutoff} \Lambda&=& \left(\frac{M_{5D}^3 R^3}{(ka)^2}\right)^{1/10} k e^{-ky} = (Mm^4)^{1/5}e^{-ky}
(ka)^5(kR)^3 \\\nonumber &<& (Mm^4)^{1/5}e^{-ky} \left(\frac{R}{a}\right)^3
\end{eqnarray}
where we have used $M_{5D}^3 = M^2 m$. The first thing to note about this expression is the appearance of the
exponential factor $e^{-ky}$. This is as one would expect in AdS, because all dimension-full quantities scale with
their position along the extra dimension, and in our case it is the effective cutoff on each brane. It is tempting to
try and raise the cutoff by taking $ka$ to be very large, but as discussed earlier the integral in equation
(\ref{eqn:AdS phi in terms of psi}) then scales as $a$ and not $1/k$. The cutoff is therefore bounded from above. In
addition, note that at least formally, the $a\rightarrow 0$ limit is healthy because while the heaviest mode in the
theory grows as $1/a$, so does the cutoff as long as we keep $ka$ fixed. This is precisely in accord with our initial
intuition. However, this is not a healthy limit because the effective theory on each site makes very little sense once
$1/a$ is greater than the 4-D Planck scale on each site. If we desire the effective cutoff on each brane to be as high
as $M e^{-ky}$, which is the quantum gravity scale, we see that there is a bound on how localized the mode can be,
\begin{equation}
 R > \left(\frac{M}{k}\right)^{8/3} a
\end{equation}
This bound might seem surprising at first as $M$ appears in the numerator, but it is simply a reflection of the fact
that the cutoff $\Lambda$ scale as $M^{1/5}$. To keep the theory healthy $k$ must be chosen appropriately so all the
modes are accommodated.

The above might leave the impression that any profile will do as long as curvature is present. However, as pointed out
earlier, it is the fact that the integral in equation (\ref{eqn: phi in terms of psi}) goes as the curvature $1/k$ that
is responsible for the healthiness of the lone wave-length modes (in contrast to the flat case where it goes like the
size of the mode). If the profile is not monotonically decreasing and has a large flat part to it this is no longer
true! When considering a profile such as shown in figure (\ref{fig: non-mono profile}) it is clear that localized modes
on either side of the well are still healthy (the integral in equation (\ref{eqn: phi in terms of psi}) still goes as
the local curvature scale). However, those modes which have support over the flat region will necessarily develop the
same ailments as modes in flat space. This is simply due to the fact that the healthy part of the kinetic term is
proportional to $\partial_y f(y)$ which vanishes.

\begin{figure}
\begin{center}
\includegraphics[scale=0.65]{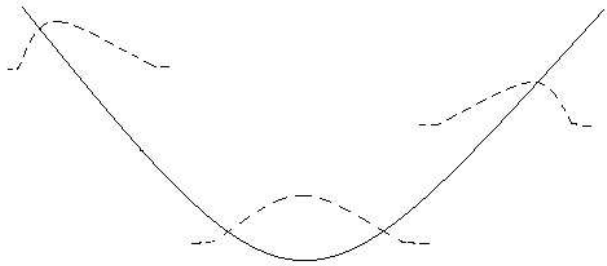}
\end{center}
\caption{A non-monotone profile. Localized modes in a curved region still have a healthy kinetic term. Modes which can
sense the flatness of the well, however, develop the same problems one encounters in the flat Moose case.} \label{fig:
non-mono profile}
\end{figure}

The above arguments were mostly qualitative and aimed for an
intuitive understanding of the prospects of discretizing AdS space.
A more quantitative analysis involving a numerical study of the
eigen-modes, their localization and interactions can be found in
\cite{discrete}
\section{Conclusions}
\label{sec: conclusions}

The ailments of a flat chain is a result of the maximal mixing between the neighboring modes which lead to strong
correlations between distant sites. The introduction of a scale difference between the sites acts as a coupling which
suppresses the mixing and renders the long wave-length theory healthy. While it is true that the strong coupling scale
is lower than expected in the continuum theory (i.e. $M \exp(-ky)$), there is a precise sense in which the effective
theory at long distances resembles that of a genuine extra dimension, albeit curved.

\textbf{Acknowledgments}: We thank Nima Arkani-Hamed and Lisa Randall for inspiring this project and providing helpful
guidance throughout this work. We gratefully acknowledge Shiyamala Thambyahpillai and Jesse Thaler for helpful
discussion.
\end{fmffile}


\begin{thebibliography}{99}

%\cite{Arkani-Hamed:2001ca}
\bibitem{Arkani-Hamed:2001ca}
  N.~Arkani-Hamed, A.~G.~Cohen and H.~Georgi,
  %``(De)constructing dimensions,''
  Phys.\ Rev.\ Lett.\  {\bf 86}, 4757 (2001)
  [arXiv:hep-th/0104005].
  %%CITATION = HEP-TH 0104005;%%

%\cite{Alishahiha:2001nb}
\bibitem{Alishahiha:2001nb}
  M.~Alishahiha,
  %``(De)constructing dimensions and non-commutative geometry,''
  Phys.\ Lett.\ B {\bf 517}, 406 (2001)
  [arXiv:hep-th/0105153].
  %%CITATION = HEP-TH 0105153;%%

%\cite{Bander:2001qk}
\bibitem{Bander:2001qk}
  M.~Bander,
  %``Gravity in dynamically generated dimensions,''
  Phys.\ Rev.\ D {\bf 64}, 105021 (2001)
  [arXiv:hep-th/0107130].
  %%CITATION = HEP-TH 0107130;%%

%\cite{Sugamoto:2001uk}
\bibitem{Sugamoto:2001uk}
  A.~Sugamoto,
  %``4d gauge theory and gravity generated from 3d ones at high energy,''
  Prog.\ Theor.\ Phys.\  {\bf 107}, 793 (2002)
  [arXiv:hep-th/0104241].
  %%CITATION = HEP-TH 0104241;%%

%\cite{Sugamoto:2002pk}
\bibitem{Sugamoto:2002pk}
  A.~Sugamoto,
  %``Generation of 4d gauge theory and gravity from their 3d versions:
  %Asymptotic disappearance of space and time (ADST) scenario,''
  Grav.\ Cosmol.\  {\bf 9}, 91 (2003)
  [arXiv:hep-th/0210235].
  %%CITATION = HEP-TH 0210235;%%

%\cite{Kan:2002rp}
\bibitem{Kan:2002rp}
  N.~Kan and K.~Shiraishi,
  %``Multi-graviton theory, a latticized dimension, and the cosmological
  %constant,''
  Class.\ Quant.\ Grav.\  {\bf 20}, 4965 (2003)
  [arXiv:gr-qc/0212113].
  %%CITATION = GR-QC 0212113;%%

%\cite{Randall:2002qr}
\bibitem{Randall:2002qr}
  L.~Randall, Y.~Shadmi and N.~Weiner,
  %``Deconstruction and gauge theories in AdS(5),''
  JHEP {\bf 0301}, 055 (2003)
  [arXiv:hep-th/0208120].
  %%CITATION = HEP-TH 0208120;%%

%\cite{Arkani-Hamed:2002sp}
\bibitem{Arkani-Hamed:2002sp}
  N.~Arkani-Hamed, H.~Georgi and M.~D.~Schwartz,
  %``Effective field theory for massive gravitons and gravity in theory space,''
  Annals Phys.\  {\bf 305}, 96 (2003)
  [arXiv:hep-th/0210184].
  %%CITATION = HEP-TH 0210184;%%

%\cite{Arkani-Hamed:2003vb}
\bibitem{Arkani-Hamed:2003vb}
  N.~Arkani-Hamed and M.~D.~Schwartz,
  %``Discrete gravitational dimensions,''
  Phys.\ Rev.\ D {\bf 69}, 104001 (2004)
  [arXiv:hep-th/0302110].
  %%CITATION = HEP-TH 0302110;%%

%\cite{Schwartz:2003vj}
\bibitem{Schwartz:2003vj}
  M.~D.~Schwartz,
  %``Constructing gravitational dimensions,''
  Phys.\ Rev.\ D {\bf 68}, 024029 (2003)
  [arXiv:hep-th/0303114].
  %%CITATION = HEP-TH 0303114;%%

%\cite{Randall:1999ee}
\bibitem{Randall:1999ee}
  L.~Randall and R.~Sundrum,
  %``A large mass hierarchy from a small extra dimension,''
  Phys.\ Rev.\ Lett.\  {\bf 83}, 3370 (1999)
  [arXiv:hep-ph/9905221].
  %%CITATION = HEP-PH 9905221;%%

%\cite{Vainshtein:1972sx}
\bibitem{Vainshtein:1972sx}
  A.~I.~Vainshtein,
  %``To The Problem Of Nonvanishing Gravitation Mass,''
  Phys.\ Lett.\ B {\bf 39}, 393 (1972).
  %%CITATION = PHLTA,B39,393;%%

%\cite{Randall: tba}
\bibitem{discrete}
  L.~Randall,, M.~D.~Schwartz, and S.~Thambyapillai
  %``Discretizing Gravity in Warped Spacetime,''
  arXiv:hep-th/0507102.
  %%CITATION = HEP-TH 0507102;%%

\end{thebibliography}
\end{document}